\documentclass[aps,twocolumn,showpacs]{revtex4}
\usepackage{graphicx}
\usepackage{dcolumn}
\usepackage{bm}

\begin{document}

\title{Transport and scaling analysis in the relativistic Standard map}

\author{Andr\'e L.\ P.\ Livorati$^{1}$, Marcelo de Almeida Presotto $^1$, João Victor Valdo Mascaro$^{1}$}

\affiliation{$^1$ São Paulo State University (UNESP), Institute of Geosciences and Exact Sciences, Rio Claro, Brazil.
\\ 
}


\begin{abstract}

We investigate some statistical and transport properties of the relativistic standard map. Through the Hamiltonian of a wave packet under an electric potential, we are able to obtain a relativistic version of the standard map, where there are two control parameters that rule the dynamics: $K$, which is the classical intensity parameter, and $\beta$, which controls the relativity. The phase space is mixed and exhibits confined local chaos for $\beta$ near unity, approaching integrability. As $\beta$ is diminished (entering the semi-classical regime), diffusion in the action variable begins to occur. However, the phase space loses its axial symmetry and an invariant curve appears to limit the diffusion as $\beta$ gets smaller. We investigate the diffusion in the action variable as a function of the number of iterations, showing that the root mean square action grows initially and bends towards a saturation regime for long times. Scaling properties were established for this behavior as a function of $\beta$, and a perfect collapse of the curves was obtained, indicating scaling invariance. Additionally, we investigated the transport properties concerning the survival probability of initial conditions. The decay rates of the survival probability are mainly exponential, followed by power-law tails. As we vary the value of $\beta$, the escape rates become slower and also obey a scaling law in their decay.
\end{abstract}
\maketitle

\section{Introduction}
\label{sec1}

Hamiltonian systems are typically non-integrable and non-ergodic, exhibiting a mixed phase space with chaotic seas, invariant tori, KAM islands, and cantori\cite{meiss,litchenberg}. This complex structure strongly affects transport properties\cite{zaslasvsky2007,zaslasvsky2008}. While strongly chaotic systems display normal diffusion with exponential escape rates, systems with mixed phase space often present anomalous transport characterized by non-exponential decay, such as power-law or stretched exponential behavior \cite{gaspard,alltman}.

The mechanism behind this phenomenon is stickiness: chaotic trajectories can remain trapped near the boundaries of stability islands for long but finite times. This intermittent behavior alternates between chaotic diffusion and quasi-regular motion, generating long-time correlations \cite{stick1, stick2, stick3, stick4, stick5, stick6,stick7,stick8,stick9,stick10,stick11}. Applications includes many research areas as: galactic dynamics\cite{astronomy1,astronomy2}, plasma physics \cite{plasma1,plasma2,plasma3,plasma4,plasma5,plasma6,plasma7}, fluid mechanics\cite{fluids}, acoustics\cite{acoustics}, and biology \cite{biology}. A common approach to study these effects is through escape-time statistics. The survival probability $\rho(n)$ consider the statistics over an orbit that does not escape through a hole up to time $n$, and is highly sensitive to the underlying dynamics. Its deviation from exponential decay provides clear evidence of stickiness and the intricate interplay between chaotic and regular regions. \cite{open1,open2}.

The model under study in this paper is the relativistic standard map. Introduced in 1989 by Chernikov, Tél, Vattay and Zaslavsky \cite{pra89}, the mapping describes de dynamics of charged particles in the field of wave packets excited a result of external perturbation of plasma by electromagnetic waves. After considering relativistic Kinetic energy, Hamilton Equations and the correction by the Lorentz factor, one can obtain the relativistic standard mapping (RSM)\cite{rsm1,rsm2,rsm3,rsm4}. Among several physical applications of the RMS we can highlight: plasma physics and magnetic fusion researches \cite{rsm_plasma}, where particle confinement and radio-frequency heating, particularly when accounting for dissipative mechanisms like synchrotron radiation, generates complex stochastic attractors and solid-state physics \cite{rsm_solid}, through the relativistic dissipative Frenkel-Kontorova model, which characterizes the nonlinear dynamics of thermoelectric materials and spatially modulated incommensurate lattice structures \cite{rsm_frenkel1,rsm_frenkel2}. Also, one may find applications of the dissipative version of the RSM considering: Shrimp Shape domains \cite{rsm_shrimp}, Boundary crisis \cite{rsm_crisis1, rsm_crisis2} and many attractor systems \cite{rsm_attractors1,rsm_attractors2}, among others.

In this paper, we characterized the dynamics and transport properties of the relativistic standard map, focusing on scaling analysis and diffusion behavior. Our results demonstrate that the phase space is mixed, and the diffusion in the action variable is bounded by invariant spanning curves, leading to a saturation regime for the root mean square action at long times. We established that this diffusion process obeys scaling laws, allowing for a universal collapse of the $I_{RMS}$ curves. Through the analysis of survival probability, we found that the escape of orbits follows an initial exponential decay followed by power-law tails, which are signature of stickiness near stability islands in the phase space. Furthermore, the exponential decay rate was shown to be scaling invariant when properly rescaled. Our investigation of the escape basins and escape angles also highlighted the fundamental role of stable and unstable manifolds in creating preferential escape routes. These findings provide a comprehensive description of transport in relativistic Hamiltonian systems and confirm the existence of universal scaling properties despite the complexity introduced by relativity.


The paper is organized as follows: In Sec. \ref{sec2} we describe the details of the RSM mapping and some chaotic properties. Section \ref{sec3} is devoted to the statistical
analysis of the $(I_{rms})$ curves, as well as, the investigation of the transport and diffusion, concerning escape basins, survival probability curves and the scaling analysis. Finally, in
Sec.\ref{sec4} we drawn some final remarks, conclusions and perspectives.

\section{The model, the mapping and chaotic properties}
\label{sec2}
In this section, we will describe the mapping that leads to the Relativistic Standard Mapping. Let us start with the original standard mapping \cite{Chirikov}, where derived from the $\delta-$kicked rotator, one can obtain the following Hamiltonian

\begin{equation}
H = \frac{J^{2}}{2} + K \cos(\theta) \sum_{n=-\infty}^{\infty} \delta\left(\frac{t}{T} - n\right),
\label{eq1}
\end{equation}
where $K$ is the amplitude of the perturbation delta function (kicks) with period $T=2\pi/\nu$. The equation of motion, found from the integration of Hamilton's equations, is given by:
\begin{equation}
\left\{\begin{array}{ll}
\dot{J}=K \sin (\theta) \sum_{n=-\infty}^{\infty} \delta\left(\frac{t}{T}-n\right), \\
\dot{\theta}=J
\end{array}\right.
\label{eq2}
\end{equation}
Assuming that $(J_{n}, \theta_{n})$ are the values of the variables immediately before the $n$-th kick, and that $(J_{n+1}, \theta_{n+1})$ represent their values immediately before the $(n+1)$-th kick, the standard mapping can be written as:
\begin{equation}
S:\left\{\begin{array}{l}
J_{n+1}= J_{n}+K \sin \left(\theta_{n}\right) \\
\theta_{n+1}=\theta_{n}+J_{n+1} \quad \bmod (2 \pi)
\end{array}\right.
\label{eq3}
\end{equation}

\subsection{Relativistic Standard Map}

Let us now consider the particle model for a wave packet, introduced in the literature by G. Zaslavsky et al., \cite{pra89,rsm2,rsm3} where the electric field $E(x,t)$ is written as:
\begin{equation}
E(x,t) = \sum_{n=-\infty}^{\infty} E_{n} \sin(k_{n}x - \omega_{n}t),
\label{eq4}
\end{equation}
where $E_{n}$ is the amplitude of the $n$-th Fourier component of the electric field wave. Considering that the wave packet has a broad spectrum, we can assume that $E_{n}=E_{0}$, $k_{n}=k_{0}$, and $\omega_{n}=n\omega$. Thus, we find:
\begin{equation}
E(x,t) = E_{0} \sin(k_{0}x) \sum_{n=-\infty}^{\infty} \cos(n\omega t),
\label{eq5}
\end{equation}
and using the Fourier decomposition on the periodic Dirac delta function, we find:
\begin{equation}
E(x,t) = E_{0}T \sin(k_{0}x) \sum_{n=-\infty}^{\infty} \delta(t - nT),
\label{eq6}
\end{equation}
where $T=2\pi/\omega$.

Assuming that the motion of an electron, with rest mass $m_{0}$ and charge $-e$ in an electric field given by Equation (\ref{eq6}), can be described by the relativistic Hamiltonian:
\begin{equation}
H(x,p,t) = \sqrt{p^{2}c^{2} + m_{0}^{2}c^{4}} - \frac{eE_{0}T}{k_{0}} \cos(k_{0}x) \sum_{n=-\infty}^{\infty} \delta(t - nT),
\label{eq7}
\end{equation}
where $c$ is the speed of light and $p = m_{0}v/\sqrt{1-(v/c)^{2}}$ is the relativistic momentum. Thus, between two consecutive kicks, the particle is in free motion. In this way, we can find a system of two differential equations from the Hamiltonian of Eq. (\ref{eq7}):
\begin{equation}
\left\{\begin{array}{ll}
\dot{x} = \frac{pc^{2}}{\sqrt{p^{2}c^{2} + m_{0}^{2}c^{4}}}  \\
\dot{p} = -eE_{0}T \sin(k_{0}x) \sum_{n=-\infty}^{\infty} \delta(t - nT).
\end{array}\right.
\label{eq8}
\end{equation}
Defining $\theta = k_{0}x$ and introducing the auxiliary variables $\beta = \omega/kc$ and $I = k_{0}p/m_{0}\omega$, and assuming that $(I_{n}, \theta_{n})$ are the values of the variables immediately before the $n$-th kick, and that $(I_{n+1}, \theta_{n+1})$ represent the values immediately before the $(n+1)$-th kick, we introduce the relativistic standard map:
\begin{equation}
S_{rel}: \left\{\begin{array}{ll}
I_{n+1} = I_{n} + K \sin(\theta_{n}) \\
\theta_{n+1} = \theta_{n} + \frac{I_{n+1}}{\sqrt{1+(\beta I_{n+1})^{2}}} \pmod{2\pi}
\end{array}\right.
\label{eq9}
\end{equation}
where $K = 2\pi e E_{0} k_{0} / m_{0} \omega^{2}$ is the parameter that controls the transition from integrability ($K=0$) to non-integrability ($K \neq 0$). In the limit $\beta \to 0$, the relativistic standard map reduces to the Chirikov standard map, given in Eq. (\ref{eq3}). In the ultra-relativistic limit $\beta \to \infty$ or $I_{n} \to \infty$, it is observed that the dynamics tends towards integrability.

\begin{figure}[htb]
\begin{center}
\centerline{\includegraphics[width=9cm,height=12cm]{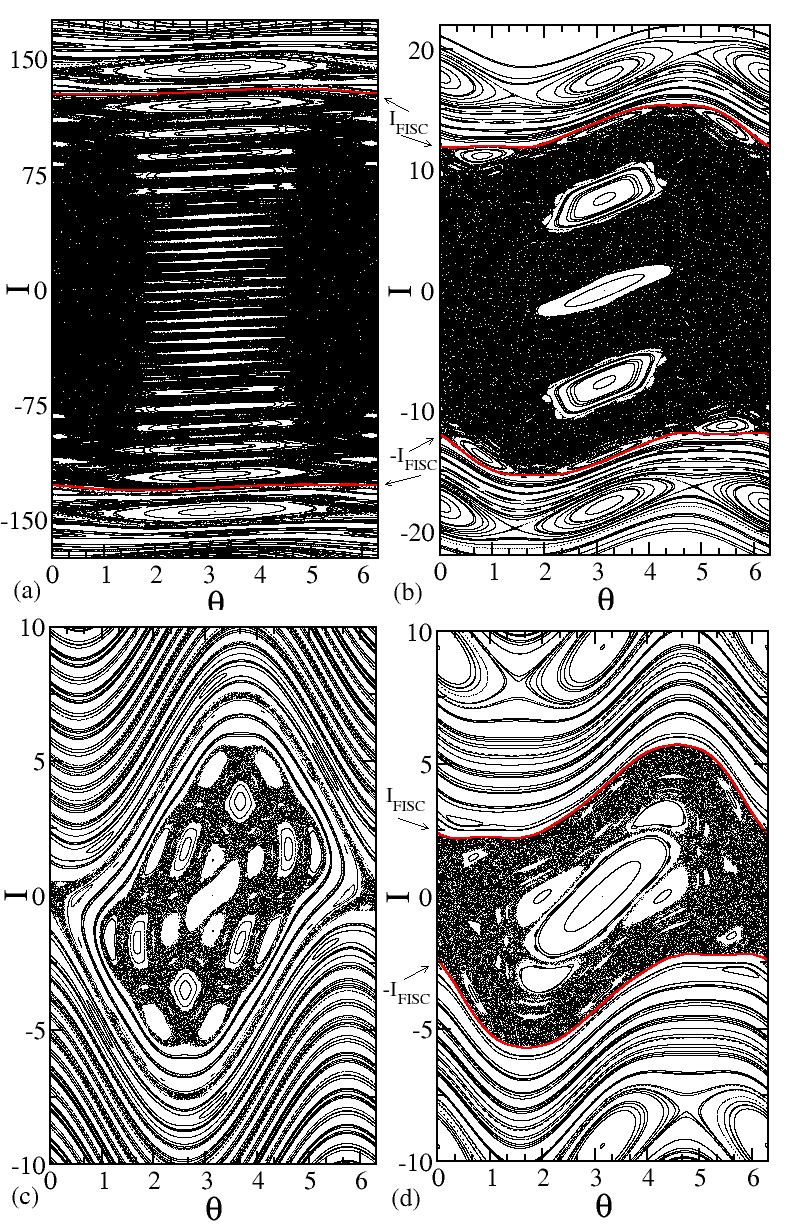}}
\end{center}
\caption{Color online: {\it Phase space for the complete dynamics of the RSM. In (a) $\beta=0.009$, in (b) $\beta=0.1$, in (c) $\beta=0.9$ and in (d) $\beta=0.3$. Also, we have depicted (in red) the position of the first invariant spanning curve $(I_{FISC})$, concerning upper positive and lower negative edges. In all items the control parameter was kept fixed as $K=3.5$}}
\label{fig1}
\end{figure}

Figure \ref{fig1} shows the phase space for four different values of $\beta$, for a fixed $K=3.5$ for $50$ different initial conditions. One can see that the phase space presents basically a mixed structure for all values of $\beta$. They have a chaotic sea embedded by two invariant spanning curves in the upper and lower regimes, displayed as $\pm I_{FISC}$. Also, there is a chain of islands appearing as the action $I$ is increased (decreased for negative values), and the number of the islands tends to grow as the closer we get to the $\pm I_{FISC}$ curves, indicating a symmetry break along the chaotic sea, if one consider juts positive (or negative) portion of the phase space. In particular, for Fig.\ref{fig1}(c) where $\beta=0.9$, the chaotic sea is confined to a small portion of the phase space, not allowing a diffusion along the action axis. Such diffusion only happens when $\beta$ diminishes, as one can see in Figs.\ref{fig1}(a,b,d), where we have respectively $\beta=0.009, \beta=0.01$ and $\beta=0.3$.

Through the analysis of the phase space displayed in Fig.\ref{fig1} we can visualize that the introduction of the relativistic parameter $\beta$, affects globally the shape of the phase space, and prevents the unlimited diffusion in the action variable, as invariant curves appears in the upper and lower regimes. The position of the $\pm I_{FISC}$ varies with $\beta$, and an analytical approach for its position can be set as follows.

\subsection{Characterization of the first invariant curve}

We suppose that near an invariant tori, which limits the size of the chaotic sea, $I$ can be
written as \cite{invar1,invar2}
\begin{equation}
I_{n+1}\approx I^*+\delta I_{n+1}~,
\label{eq10}
\end{equation}
where $I^*$ is a typical value of the invariant curve and $\delta I_{n+1}$ is a small perturbation of $I_{n+1}$.
Now the first equation of mapping (\ref{eq9}) can be written as $ \delta I_{n+1}=\delta I_n+K\sin{\theta_n}$. Using then the second equation of mapping (\ref{eq9}, we have:
\begin{equation}
\begin{array}{ll}
\theta_{n+1}=\theta_n+{{I^*+\delta I_{n+1}}\over{{\sqrt{1+(\beta (I^*+\delta I_{n+1}))^{2}}}}}~,\\
=\theta_n+I^*\left(1+{\delta I_{n+1}\over I^*}\right)\left[1+\beta^2 {I^*}^2\left(1+{\delta I_{n+1} \over I^*}   \right)^2   \right]^{-1/2}~.
\end{array}
\label{eq11}
\end{equation}
Expanding the Eq.(\ref{eq11}) into Taylor series around $\delta I_{n+1}/I^*$, keeping only terms
of ﬁrst order, we have
\begin{equation}
\theta_{n+1}=\theta_n+{I^* \over \sqrt{1+\beta^2{I^*}^2}}+{\delta I_{n+1} \over {(1+\beta^2{I^*}^2)^{3/2}}}~.
\label{eq12}
\end{equation}
Defining the new variable $I_{n+1}$ as
\begin{equation}
I_{n+1}= {I^* \over \sqrt{1+\beta^2{I^*}^2}}+{\delta I_{n+1} \over {(1+\beta^2{I^*}^2)^{3/2}}}~
\label{eq13}
\end{equation}
we find that $\theta_{n+1}=\theta_n + I_{n+1}$. Doing the appropriate algebra we end up with
\begin{equation}
\begin{array}{ll}
{\delta I_{n+1} \over {(1+\beta^2{I^*}^2)^{3/2}}}+{I^* \over \sqrt{1+\beta^2{I^*}^2}}={\delta I_{n} \over {(1+\beta^2{I^*}^2)^{3/2}}}+~\\
{I^* \over \sqrt{1+\beta^2{I^*}^2}}+K\sin(\theta_n)~.
\end{array}
\label{eq14}
\end{equation}
Comparing the result in Eq.(\ref{eq14}) with the SM given in Eq.(\ref{eq3}), considering the transition from local to global chaos, when $K=K_{eff}=0.9716...$, one can set that
\begin{equation}
K_{eff}={K \over {(1+\beta^2 {I^*}^2)^{3/2}}}~.
 \label{eq15}
\end{equation}
Isolating $I^*$ from Eq.(\ref{eq15}), we can make an estimative of the position of the first invariant curve as function of the control parameters $K$ and $\beta$ for the RSM, as follows:
\begin{equation}
I^* = {1 \over \beta} \left[ \left(  {K\over K_{eff}}  \right)^{2/3} -1\right]^{1/2}~.
 \label{eq16}
\end{equation}
This result confirms the structure the phase space observed in Fig.\ref{fig1}. The position of $I^*$ is inversely proportional to the value of $\beta$, i. e., smaller the value of $\beta$ is, more accessible the phase space is for chaotic orbits to diffuse.

\section{Results and Discussion}
\label{sec3}
~~~~~In this section, we made a statistical and transport analysis for the dynamics of the RSM. Considering average properties of the quadratic action, we investigated the diffusion along the accessible phase space, and set up a transition from growth to saturation regime. Such transition was characterized by scaling arguments and an universal plot was obtained to give robustness to the scaling hypotheses.
We also calculate the decay of correlation times, also known as, the survival probability, for an ensemble in the symmetry action near zero, inside the limit region in the phase space. Stickiness shows itself inherent in the system, and affects the survival probability. At last, but not the least, we set a scaling analysis for the survival probability, considering the exponential decay rate.

\subsection{Numerical statistical analysis for the action}
Let us start by evaluating numerically the behavior of the root mean square action, which is made by considering $I_{RMS}=\sqrt{\overline{I^2}}$, where
\begin{equation}
\overline{I^2}={1\over M} \sum_{i=1}^M {1\over n} \sum_{j=1}^n {(I_{i,j})}^2~,
\label{eq17}
\end{equation}
where $M$ is the ensemble of initial conditions, and $n$ is the number of iterations. The average is taken along the orbit and along
the ensemble of initial conditions. The initial conditions where always chosen in the chaotic sea, uniformly distributed along the phase $\theta\in[0,2\pi)$,
considering a regime where the initial action $I_0\approx\pm\beta$. Such scenario of initial conditions, guarantees an uniform diffusion in both positive and negative axis. Here, we took an extra care to not chose an initial condition that would lie inside a stability islands, otherwise it would damage the statistics.

\begin{figure}[h!]
\begin{center}
\centerline{\includegraphics[width=9.5cm,height=8cm]{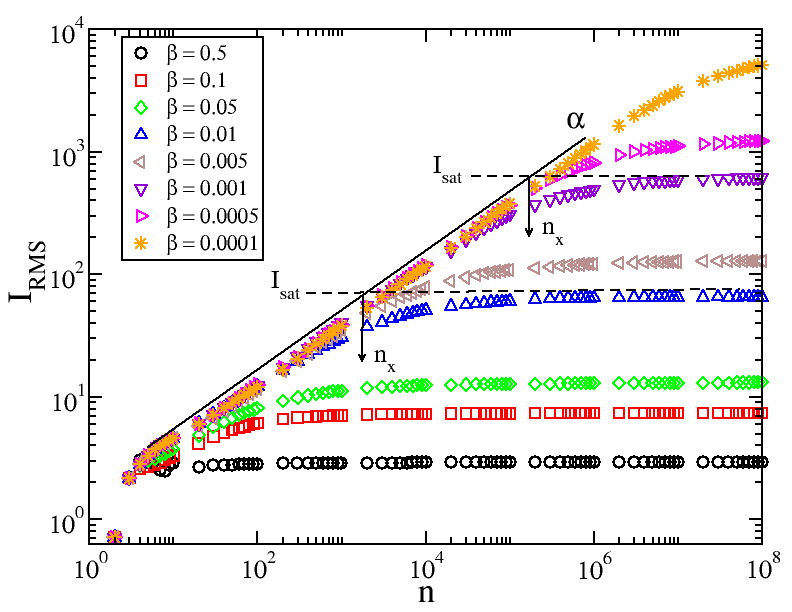}}
\end{center}
\caption{{\it $I_{RMS}$ curves of the RSM, for different values of $\beta$, as the parameter $K=3.5$ was kept constant. One can see that the $I_{RMS}$ present a growing regime according to $\alpha\approx 0.5$ exponent, then they suffer a crossover $(n_x)$ and evolve to a saturation plateau $(I_{sat})$ for long iteration times. As expected by Eq.(\ref{eq16}), the saturation plateaus are inversely proportional to the value of $\beta$.}}
\label{fig2}
\end{figure}

Also, it is important to clarify that if we had took a simple average over the action, there would be no meaningful statistics to study, since the average would be around zero, once the phase space has accessible regions in both positive and negative axis, in a symmetric manner. That is why we considered the average over the root mean square action, given by Eq.(\ref{eq17}).

Figure \ref{fig2} displays the $I_{RMS}$ curves evaluated over an ensemble of $5000$ initial conditions,
iterated up to $10^8$. We have considered at least, three decades along side the parameter $\beta$. The parameter $K=3.5$ was kept fixed. As already ``foreseen'' by the phase space, shown in Fig.\ref{fig1} and by Eq.(\ref{eq16}), as smaller the value of $\beta$, higher is the diffusion in the action axis. One may observe in Fig.\ref{fig2}, that all curves of $I_{RMS}$ start
in a growth regime, according an $\alpha\approx1/2$ exponent, and suddenly they bend towards a crossover $(n_x)$, and pass to a steady state plateau, in a saturation regime, given by $I_{sat}$.

\subsection{Scaling Analysis}
One can see, that all curves obeys the same behavior, a growing regime for short time, passing through a crossover and then a saturation plateau for long times. And, since the parameter $K=3.5$ was kept constant, we
propose some scaling hypothesis depending on $\beta$ to describe the $I_{RMS}$ curves.

\begin{itemize}

\item
{\it(i)} $I_{RMS}\propto n^{\alpha}$, when we have $n\ll n_x$.\

\item
{\it(ii)} $I_{sat}\propto \beta^{\delta}$, for $n\gg n_x$.\

\item
{\it(iii)} $n_x\propto \beta^{\nu}$,   where $n_x$ is the crossover iteration.\

\end{itemize}

\begin{figure}[h!]
\begin{center}
\centerline{\includegraphics[width=9cm,height=10cm]{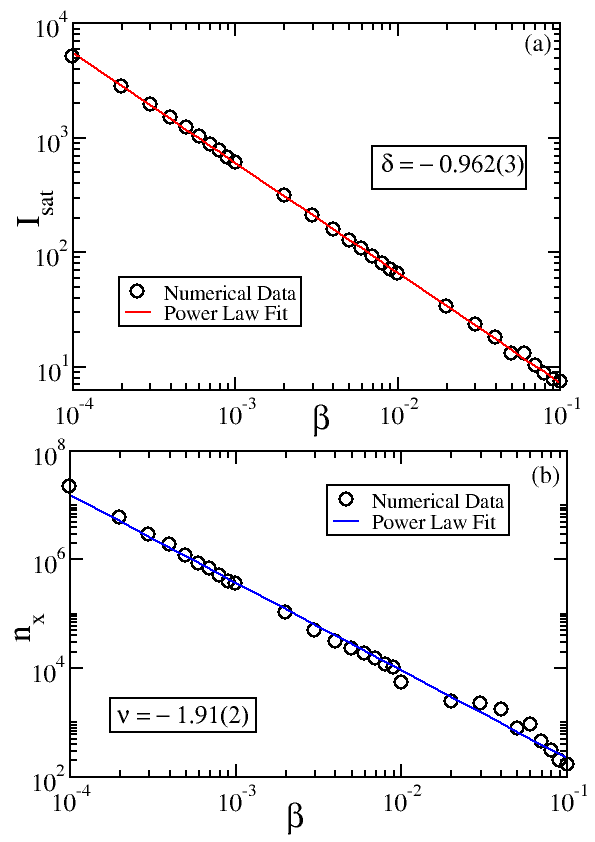}}
\end{center}
\caption{Color online: {\it Critical exponents obtained by numerical simulations. In (a) plot of $I_{sat}\times\beta$ furnishing us $\delta=-0.962(3)$,
in (b) there is a plot of $n_x\times\beta$, where we obtain $\nu=-1.91(2)$.}}
\label{fig3}
\end{figure}

It is shown in Fig.\ref{fig3} a numerical analysis of both saturation and crossover regimes. In Fig.\ref{fig3}(a) a plot $I_{sat}\times \beta$, furnishing us a power law fit of $\delta=-0.962(3)$. And in Fig.\ref{fig3}(b), one can obtain the exponent $\nu=-1.91(2)$, through a power law fit in a plot of $n_x \times \beta$. All the results shown in Fig. \ref{fig3} were obtained using very long simulations of $10^{8}$ iterations. The ensemble average
used was $M=5\times 10^3$.

Using now the formalism of describing the scaling analysis with a homogeneous generalized function \cite{stick3,stick4,stick6}, we may obtain the following expression
\begin{equation}
I_{RMS}(n,\beta)=\lambda I_{RMS}(\lambda^a n,\lambda^b\beta)~,
\label{eq18}
\end{equation}
where, $\lambda$ is a scaling factor, $a$ and $b$ are scaling exponents.\

Assuming that $\lambda^a n=1$, one can obtain the following
\begin{equation}
\lambda=(n)^{-1/a}~.
\label{eq19}
\end{equation}

Replacing Eq.(\ref{eq19}) in (\ref{eq18}), we obtain
\begin{equation}
I_{RMS}(n,\beta)=n^{-1/a}{I^1_{RMS}}(1,\lambda^{-b/a}\beta)~,
\label{eq20}
\end{equation}
where ${I^1_{RMS}}$ is constant for short time, i. e., $n\ll n_x$.\

If we compare Eq.(\ref{eq20}) with the scaling hypothesis {\it(i)}, we obtain
\begin{equation}
\alpha=-{1\over a}~,
\label{eq21}
\end{equation}
and given that the critical exponent $\alpha\approx0.5$, obtained by
fitting a power law to the growing regime of the $I_{RMS}$ curves, we have that $a=-2$.\

Choosing now $\lambda^b\beta$ as a constant, we have
\begin{equation}
\lambda=\beta^{-1/b}~.
\label{eq22}
\end{equation}

Substituting Eq.(\ref{eq22}) in (\ref{eq18}), we obtain
\begin{equation}
I_{RMS}(n,\beta)=\beta^{-1/b}{I^2_{RMS}}(\lambda^{-a/b}n,1)~,
\label{eq23}
\end{equation}
where ${I^2_{RMS}}$ is assumed to be constant for $n_x\ll n$, which means the saturation regime.
A comparison of Eq.(\ref{eq23}) with hypothesis {\it(ii)} leads to
\begin{equation}
\delta=-{1\over b}~,
\label{eq24}
\end{equation}
and given that we already known the value of $\delta\approx-0.962$, from the numerical results given in Fig.\ref{fig3}(a); thus yielding $b=1.039$.\

\begin{figure}[h!]
\begin{center}
\centerline{\includegraphics[width=9cm,height=14cm]{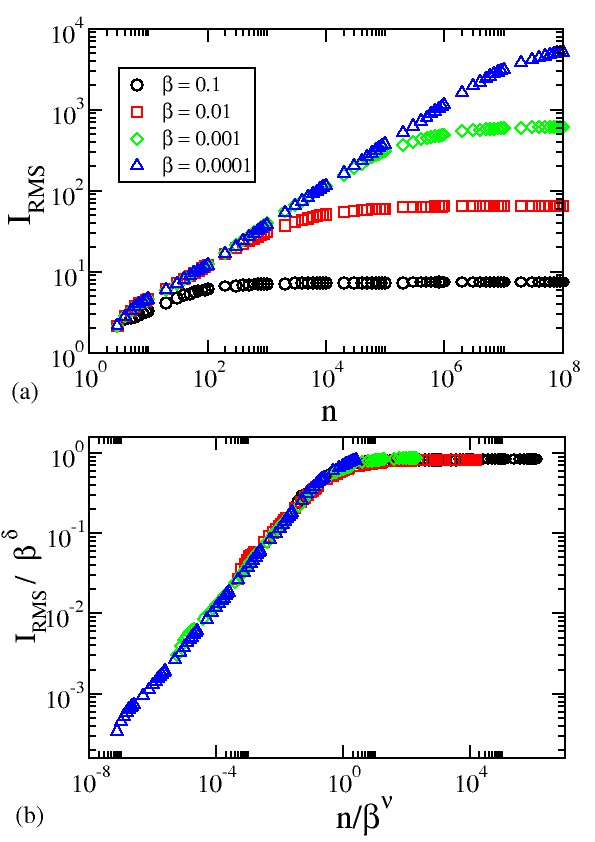}}
\end{center}
\caption{Color online: {\it In (a) we have a few curves of $I_{RMS}$ for a wide range of control parameter
$\beta$, and in (b) we show the collapse of all curves of $I_{RMS}$ displayed in (a), confirming the validity of the scaling hypothesis and the critical
exponents obtained.}}
\label{fig4}
\end{figure}

We now have two distinct values for $\lambda$, as shows Eqs.(\ref{eq19}) and (\ref{eq22}). Making a comparison
between then, we may have the following
\begin{equation}
(n)^{-1/a}~=~\beta^{-1/b}~.
\label{eq25}
\end{equation}

Considering the both sides of Eq.(\ref{eq25}), taking the power law of ${-a}$, and knowing the previously result from Eqs.(\ref{eq21}) and (\ref{eq24}) we get
\begin{equation}
n=\beta^{\delta/\alpha}~.
\label{eq26}
\end{equation}

\begin{figure*}[htb]
\begin{center}
\centerline{\includegraphics[width=18cm,height=8cm]{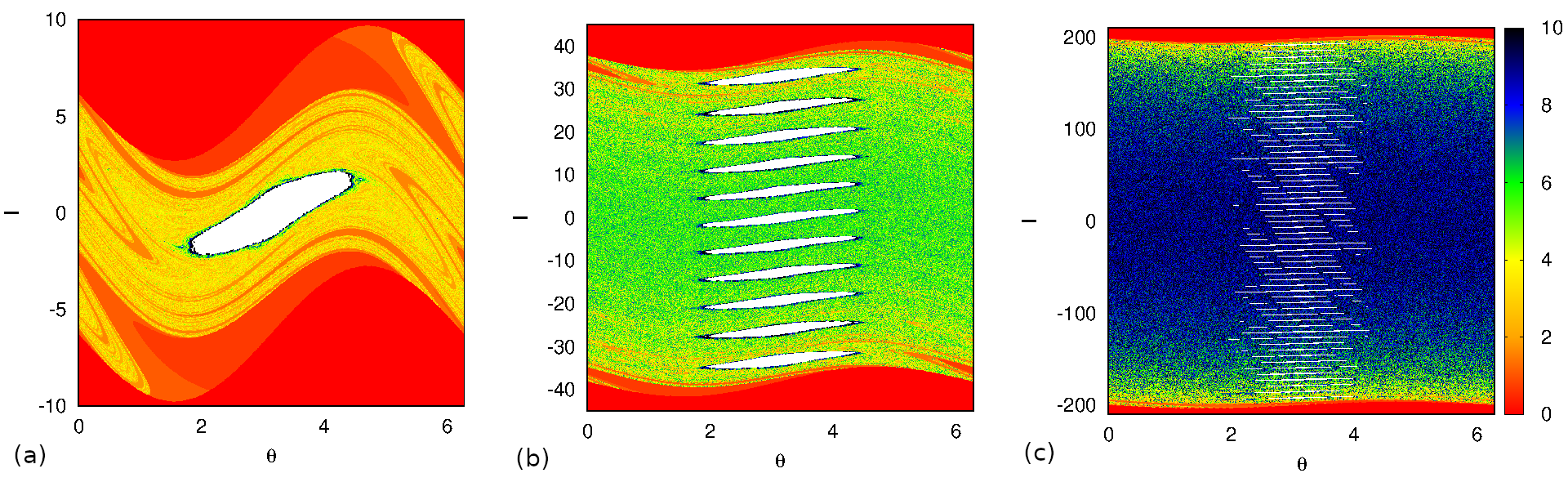}}
\end{center}
\caption{Color online: {\it Plot of the escape basins, until the grid of initial conditions reach the escape plateaus, located at the edges (positive and negative). In (a) we have $\beta= 0.1$, in (b)$\beta= 0.01$, and in (c) $\beta= 0.001$. Dark blue (black) indicates long time evolution until reaching the hole, while red (gray) indicates fast escape. White denotes the particle never
escaping until $10^8$ collisions. Particularly one can see that stickiness delays the escape influencing in the transport.}}
\label{fig5}
\end{figure*}

After a comparison with Eq.(\ref{eq26}), and scaling hypothesis {\it(iii)}, we found
\begin{equation}
\nu={\delta\over\alpha}~.
\label{eq27}
\end{equation}

Once $\alpha\approx0.5$, $\delta\approx-0.962$, one can obtain $\nu=-1.924$, which is a very close value to the one obtained numerically in Fig.\ref{fig3}(b).

Equation (\ref{eq27}) furnishes the scaling law for our system considering the behavior of the $I_{RMS}$ curves.
Considering now the scaling hypothesis and the critical exponents found, we can collapse all curves of $I_{RMS}$  into a single and universal plot.

Figure \ref{fig4}
shows how this procedure is done. In Fig.\ref{fig4}(a) we have a few curves of $I_{RMS}$ for a wide range of values of $\beta$. After a rescale
in the axis as $n\rightarrow n/\beta^{\nu}$ and
$I_{RMS}\rightarrow I_{RMS}/\beta^{\delta}$, as shown in Fig.\ref{fig4}(b), an unique and universal plot appears, confirming thus the validity of the scaling laws.

\subsection{Transport}
Since the investigation of the $I_{RMS}$ curves shown us that for long times, the curves bend towards a saturation plateau, we decided to investigate how the transport occur, by creating in somehow an escape basin. We created a grid of $1000\times1000$ initial conditions equally
distributed in the whole accessible phase space until the saturation, i.e., $\theta_0\in[0,2\pi]$ and $I_0$ around $\pm I_{sat}$. Then each initial condition was evolved in time up to the limit of $10^8$ iterations or until the saturation action $I_{sat}$ were reached. In other words, we saved the iteration that the initial conditions
took to reach the convergence plateau.

Figure \ref{fig5} displays how this transport occurs for some values of the control parameter $\beta$. The color range denotes the number of iterations (plotted in logarithmic scale) that the orbit took until reaching the escape action $I_{esc}$, and it can be interpreted as red (gray) indicating fast escape, to blue or black (black) denoting long time dynamics. For instance, a color scale marked as $6$,
represents $\exp(6)$, or about $403$ iterations until that initial condition reaches the saturation plateau.
Also, white parts represent the stability islands and denote that the orbits did not reach $I_{esc}$ until $10^8$ iterations. One can also consider this dynamical evolution until the orbits reach the $I_{esc}$
plateau, as the introduction of a leakage, or a hole in the phase space \cite{gaspard,alltman}, where orbits are allowed to ``escape''. 

One can see in Fig.\ref{fig5}(a), for $\beta=0.1$, where the escape were set in $I_{esc}=\pm 10.0$, that initial conditions escape very rapidly in very fewer iterations, and for the region comprehended around $\pm 5.0$ in the action axis, the manifolds plays the role of fast escape branches. Also near the centered island, it is possible to see, delayed escape due to stickiness entrapment. For Figs.\ref{fig5}(b,c) we have respectively $\beta=0.01$ and $\beta=0.001$, the orbits took more iteration time to reach escape. Considering Fig.\ref{fig5}(b), the escape action were set in $I_{esc}=\pm40.0$, and basically the great portion of the region comprehended between then escapes at the same time, around $\exp(5)\approx150$. Again, it is possible to see stickiness influence in the escape, through the darker regions around the stability islands. Finally, for Fig.\ref{fig5}(c), the orbits took longer times to reach the escape region around $I_{esc}=\pm200.00$. The whole scenario of Fig.\ref{fig5} is in good agreement with Figs.\ref{fig1} and \ref{fig2}, where smaller the value of $\beta$, more accessible area the orbits have to diffuse in the phase space, and more time they took to reach higher values of the action variable.

\begin{figure}[h!]
\begin{center}
\centerline{\includegraphics[width=9cm,height=12cm]{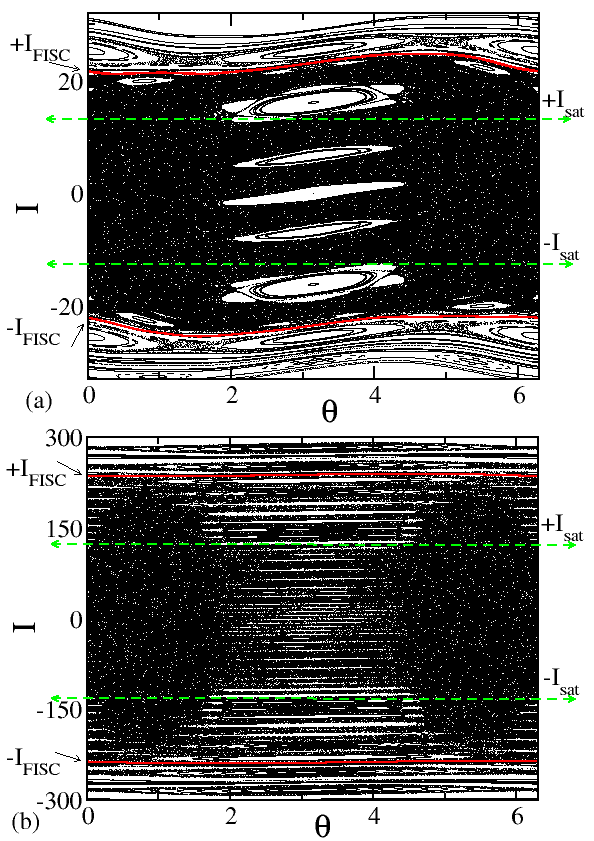}}
\end{center}
\caption{Color online: {\it Survival region of the accessible phase space comprehended between $-I_{sat},+I_{sat}$ (dashed lines) In (a) $\beta=0.05$ and in (b) $\beta=0.005$. The survival region was used to numerical evaluated the behavior of the survival probability $\rho(I,n)$.}}
\label{fig6}
\end{figure}

\subsection{Survival probability}
Since we know that the dynamics of $V_{RMS}$ will bend towards a convergence plateau for long times, we introduce two distinct holes in the action for both positive and negative regions, setup as $\pm I_{sat}$. Figure \ref{fig6} displays the how the escape regions (dashed lines) are set up according the accessible phase phase, confined between $[+I_{FISC},-I_{FISC}]$, where we have $\beta=0.05$ and $\beta=0.005$ for Figs.\ref{fig6}(a,b) respectively.

Basically, the dynamics with the introduction of a hole follows: in the positive region, we consider that an initial condition had escaped,
if its action is equal, or higher than $I_{esc}=+I_{sat}$; in the same manner, for the negative portion, an initial condition escapes if its action is equal or lower than $I_{esc}=-I_{sat}$. For both regions, we save in a vector the iteration in which the orbit had escaped, and then we build a frequency histogram for the escape, according the escape iteration.

The survival probability, described in terms of escape formalism \cite{gaspard,alltman,surv1,surv2}, is then obtained by the integration of this escape frequency histogram, as
\begin{equation}
\rho(I,n)={1\over N} {\sum_{j=1}^N} N_{rec}(n)~,
\label{eq28}
\end{equation}
where the summation is taken along an ensemble of $N=\times10^8$ initial conditions chosen along the chaotic sea and $I=I_{esc}$ is set as the escape action, or the hole position in the action axis. It is important to remark that
the initial conditions where uniformly distributed along the phase $\theta\in[0,2\pi)$,
considering a regime where the initial action $I_0\approx\pm\beta$, in a way that we can guarantee an uniform diffusion in both positive and negative axis. And, again we took an extra care to not chose an initial condition that would lie inside a stability islands. The term $N_{rec}(n)$ in Eq.\ref{eq28} denotes the number of initial conditions that did not escape through the holes until the n-th collision \cite{gaspard,alltman,surv1,surv2}.

As it is known from the literature \cite{gaspard,alltman,surv1,surv2}, the decay rate of $\rho(I,n)$ is extremely sensitive to the dynamics of the system. For strongly chaotic systems, which present
normal diffusion, the decay is typically exponential \cite{stick3,surv1,surv2}, while systems that present mixed phase space, with irregular diffusion due stickiness
influence, the decay can be slower, presenting a mix of exponential with a power law \cite{stick1, stick2, stick9}, or stretched exponential decay \cite{stick7,stick8,stick11}. For the dynamics of the RSM, which has mixed properties in the phase space, the curves of $\rho(I,n)$ may present this whole scenario of different decay rates, depending of the value of $\beta$ and the position of the hole.

\begin{figure}[ht!]
\begin{center}
\centerline{\includegraphics[width=9cm,height=8cm]{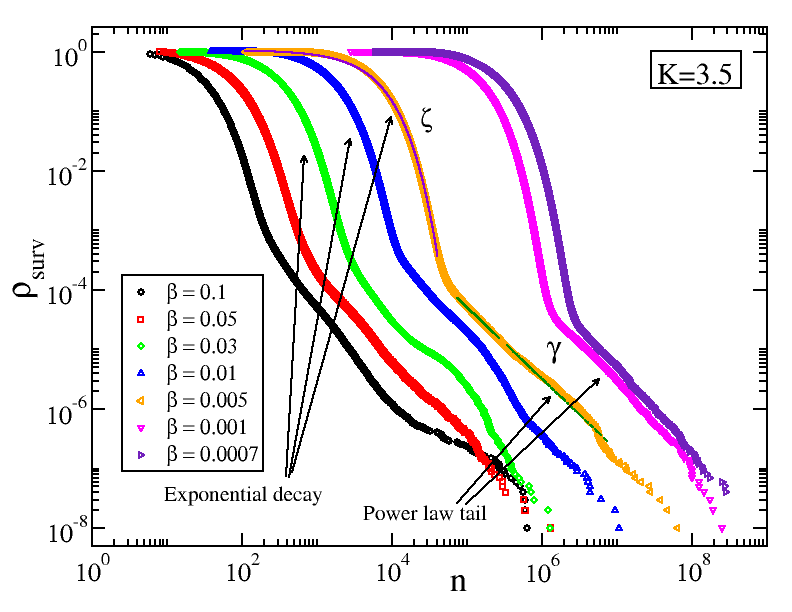}}
\end{center}
\caption{Color online: {\it Plot of the survival probability curves $\rho(I,n)$ for some values of $\beta$, where the escape action is set as $I_{esc}=\pm I_{sat}$. The curves experience an exponential decay, according $\rho(I,n)\propto e^{-\zeta n}$. By their final tails, there is a slower decay, that is due stickiness orbits. These tails are fitted according $\rho(I,n)\propto n^{-\gamma}$. Also, the we kept fixed $k=3.5$.}}
\label{fig7}
\end{figure}

Figure \ref{fig7}, shows the $\rho(I,n)$ curves for some values of $\beta$, when the escape is set in both portions of the phase space as $I_{esc}=\pm I_{sat}$, which means that an orbit could escape by the both regions (positive and negative). Basically, one can observe that the decay rate is composed of two behaviors. For short  times, it obeys an exponential decay as
\begin{equation}
\rho(I,n)=A\exp(-\zeta n)~,
\label{eq29}
\end{equation}
where $A$ is a non-negative constant and $\zeta$ is the decay rate. For long times, one can see it obeys a slower decay, that could be set as a power law according to
\begin{equation}
\rho(I,n)=Bn^{-\gamma}~,
\label{eq30}
\end{equation}
where $B$ is also a non-negative constant and $\gamma$ is the power law decay rate, that according to the literature \cite{gaspard,alltman,stick9,stick10}, is found in a range of $\gamma\in[1,3]$.

For short times, an exponential fit is in a good agreement to describe the behavior of $\rho(I,n)$, where we may obtain the decay rate $\zeta$, as shows Fig.\ref{fig7}. At the end of the exponential decay rate curve for each $\beta$, there is a changeover, and the curves of $\rho(I,n)$ starts a slower decay rate marked by the final tails
as a power law, furnishing us a $\gamma$ exponent. Such slower decay indicates an influence of some delay of theses orbits in escaping, which is due to stickiness.

\begin{figure}[ht!]
\begin{center}
\centerline{\includegraphics[width=9cm,height=12cm]{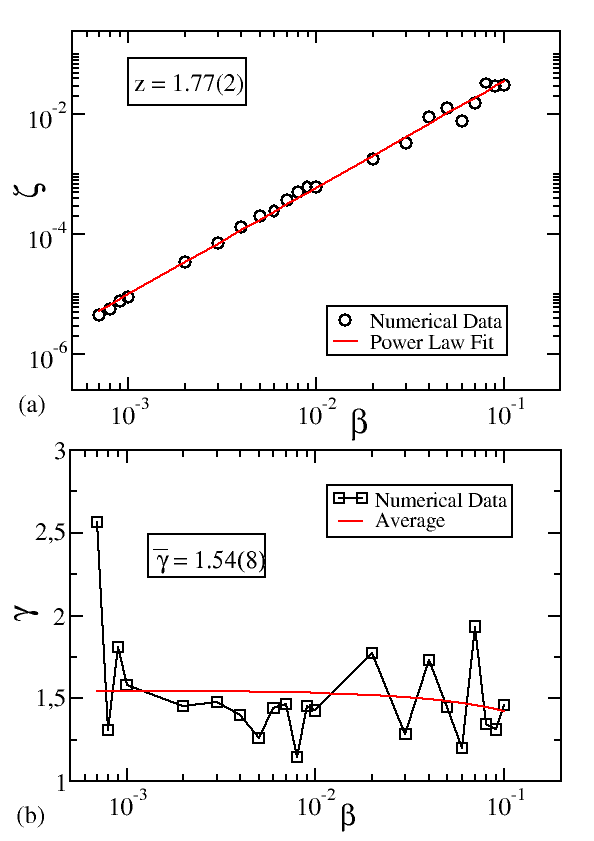}}
\end{center}
\caption{Color online: {\it In (a), we have the $\zeta$ exponent plotted as function of $\beta$. A power law fitting furnishes us a good agreement with an exponent $z=1.77(2)$. In (b). we show a plot of $\gamma \times \beta$. There is none curve fitting that shows the same mathematical beauty as observed for the exponent $z$ in (a). A simple average of the values gives us $\overline{\gamma}$=1.54(8).}}
\label{fig8}
\end{figure}

Considering both decay rates dispĺayed in Fig.\ref{fig7}, one could ask how the exponents $\zeta$ and $\gamma$ are connected with the variation of the $\beta$ parameter. Figure \ref{fig8}(a) shows a plot of the $\zeta \times \beta$, where a power law fit giver us a exponent $z=1.77(2)$, indicating that the escape rate gets slower when there is more accessible regions in the phase space for the orbits to diffuse. Also, in Fig.\ref{fig8}(b) a plot of $\gamma\times\beta$ does not seem to lead to any fitting. The value of $\gamma$ is inside the range considered to be acceptable in the literature \cite{gaspard,alltman,stick9,stick10}, and also indicates stickiness influence in the dynamics. An average over the values gives us $\overline\gamma=1.54(8)$.

\begin{figure}[ht!]
\begin{center}
\centerline{\includegraphics[width=9cm,height=12cm]{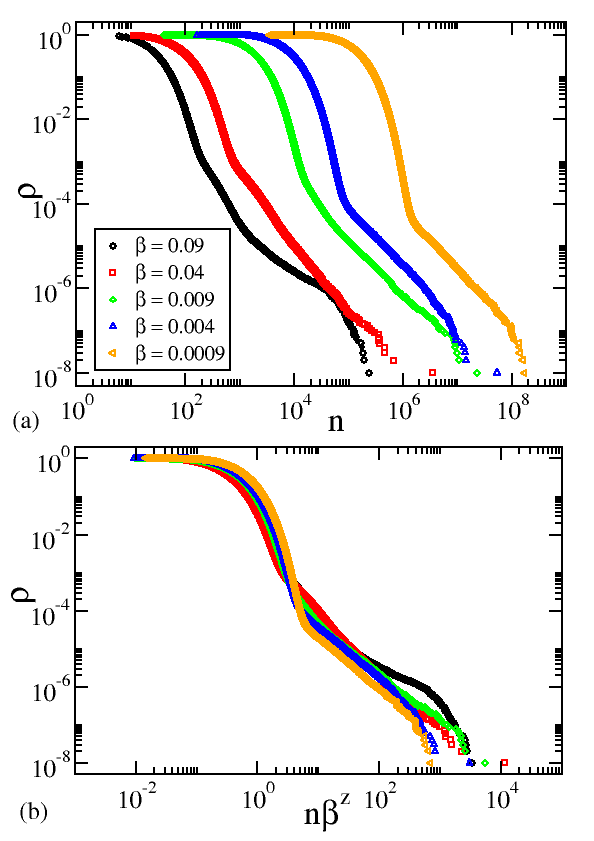}}
\end{center}
\caption{Color online: {\it In (a) we have a few curves of $\rho(I,n)$ for some values of $\beta$.
And in (b) we display a scaling invariance of the survival probability curves and its good overlap regarding the $z$ exponent obtained in Fig.\ref{fig8}(a).}}
\label{fig9}
\end{figure}

Since there is a correlation between $\beta$ and $z$ exponent, one may rearrange the horizontal axis and check if there is any scaling between then. Figure \ref{fig9}(a) shows a few curves of $\rho(I,n)$, considering a wide range of $\beta$, and in Fig.\ref{fig9}(b) we rearrange the horizontal axis as $n\rightarrow n\beta^z$, and obtain a good overlap of all $\rho(I,n)$ curves displayed in Fig.\ref{fig8}(a) indicating that the exponential decay rate is scaling invariant. Of course the scaling does no fit for the power law decays and the $\gamma$ exponent, since that they might be produced by different chains and islands shapes as we range $\beta$.

Also, one could ask about the nature of non-overlap concerning the exponent $\gamma$  and its relation
with the parameter $\beta$. In fact, we cannot assume for sure its dependence \cite{alltman,stick9}. Our supposition, is that such dependence has to do with the fractal of the system, and maybe the configuration and amount of stability islands in the accessible region for the particle. Another remarkable point lies over the manifolds, since they play a role under rapid escape \cite{stick11}, where they seem to drawn a preferential path for the orbits to escape.

\begin{figure}[ht!]
\begin{center}
\centerline{\includegraphics[width=9cm,height=12cm]{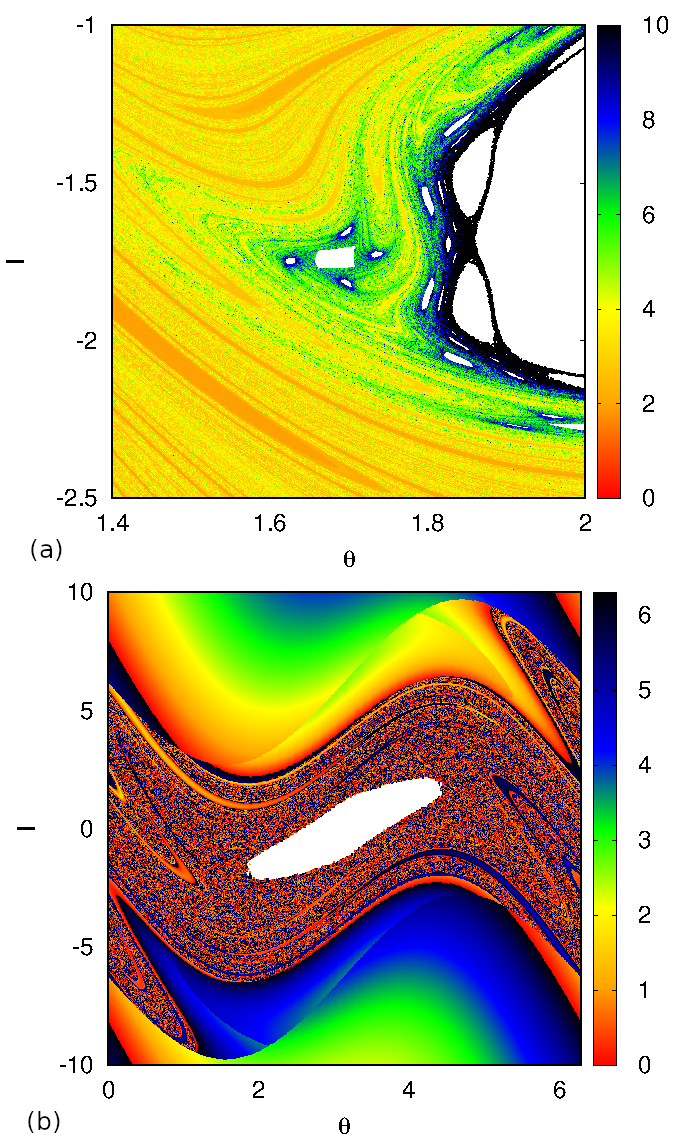}}
\end{center}
\caption{Color online: {\it In (a) we have a zoom-in window for the escape basin concerning $\beta=0.1$, where one can see the embedded scenario between the manifolds that drawn boundaries of fast and slow escape. In (b) we display the escape basin concerning the angle of escape in color palette.}}
\label{fig10}
\end{figure}

Figure \ref{fig10}(a) shows a zoom-in window for the escape basin considering $\beta=0.1$, where the color pattern palette is the same used to draw Fig.\ref{fig5}. One can see, that the manifolds in somehow drawn a boundary line between fast, midterm and slower escapes. In particular, for regions very near the chain of cantori, these regions of faster and slower escapes, are extremely embedded among themselves, creating a rich and complex escape scenario.

Basically, an orbit evolving into the strong stickiness regime layer, may leave in somehow the region delimited by these layers, and reach one of the manifold branches, thus leading the orbit to escape. So, the manifolds seems to play a crucial role, in which orbit would escape faster or slower, as already shown in other Hamiltonian mappings \cite{stick1,stick2,stick3,plasma4,plasma6,plasma7}

Another interesting fact about the escape rate, is the angle $\theta_n$ which as orbit has when we considered it had escaped. Figure \ref{fig10}(b), shows as escape basin, considering the angle of escape in the color gradient, between $[0,2\pi]$, instead of the usual iteration time. The escape here was considered as the same manner of Fig.\ref{fig5}, once an orbit reached the action in edge (positive or negative), we considered it had escaped, and then we mark the angle that the escaped orbit had, when it reached the edge. One can still see, the manifolds pattern, drawing preferential escape routes for the orbits. The interesting thing here, is that the escape angle is not uniformly distributed. If we look the accessible region for diffusion to occur, there is a preferential ``extreme angles'' in the central region, i.e., angles located between $[0,1]$ and $[5,2\pi]$. On the other hand, for the regions. near the edges, there is a better distribution of the escape angle.

Our best guess is that maybe the number of crossings between the manifolds (like a homoclinic tangle), would somehow influence the transport, producing rapid and slow escape rates depending on the number of crossings, including influencing the angle of escape, concerning preferential branches and boundaries. However, this supposition still needs further investigation and remain as an open issue.

\section{Final Remarks and Conclusions} 
\label{sec4}

To summarize, we have investigated the dynamics of the relativistic standard mapping (RSM), through the Hamiltonian point of view analysis of a charged particle under an electric field.
A nonlinear mapping was obtained and a mixed phase space was characterized composed by a chaotic sea, KAM islands and an upper/lower first invariant spanning curve $(\pm I_{FISC})$, that works as a barrier, preventing the chaotic sea to grow. Also, an analytical approach was made concerning the position of these curves.

Statistical analysis for an ensemble of initial conditions leads the root mean square action to a steady state plateau for long times. A scaling analysis was proposed, where critical exponents were numerically obtained and a scaling law was applied in order to collapse all the $I_{RMS}$ curves.  Evaluating an investigation of the survival probability, the ensemble presents typical exponential decay rates, followed by power law tails,
which mark a direct influence of sticky orbits in the dynamics. The exponential decay rates are scaling invariant with respect to the control parameter $\beta$. We also stress out that manifolds play a crucial
role in the dynamics, outlining an escape path for rapid escape, influencing which orbit would escape faster or slower, where their boundaries are embedded among manifold branches, creating a non-uniform distribution in the angle of escape.

In the near future it would be interesting to investigate the crucial role of the manifolds, from the analytical and numerical point of view, focusing in the number of crossings between them.
As well, a study of the slower decay rates, and its relation with the preferential chains of islands that orbits may prefer to stay trapped seems promising.

\acknowledgments
ALPL acknowledges FAPESP, CNPq and CAPES for financial support. MAP thanks CNPq and JVVM thanks Brazilian agency CAPES. This research was supported by resources supplied by the Center for Scientific Computing (NCC/GridUNESP) of the S\~ao Paulo State University (UNESP).

\end{document}